# Fractional Skyrmion tubes in Chiral-Interfaced Three-Dimensional Magnetic Nanowires

J Fullerton[1,2]*, N Leo[3,4]**, J Jurczyk[3], C Donnelly[5,6], D Sanz-Hernández[7], L Skoric[8], N Mille[9], S Stanescu[9], D A MacLaren[2], R Belkhou[9], A Hierro-Rodriguez[10,11] and A Fernández-Pacheco[2,3]***

[1] Materials Science Division, Argonne National Laboratory, Illinois, USA

[2] SUPA, School of Physics and Astronomy, University of Glasgow, Glasgow, UK

[3] Institute of Applied Physics, TU Wien, Vienna, Austria

[4] Department of Physics, School of Science, University of Loughborough, Loughborough, UK

[5] Max Planck Institute for Chemical Physics of Solids, Dresden, Germany

[6] International Institute for Sustainability with Knotted Chiral Meta Matter (WPI-SKCM2), Hiroshima University, Hiroshima, Japan

[7] Laboratoire Albert Fert, CNRS, Thales, Université Paris-Saclay, Palaiseau, France

[8] University of Cambridge, Cambridge, United Kingdom

[9] Synchrotron SOLEIL, Saint-Aubin, France

[10] Depto. Fisica, Universidad de Oviedo, Oviedo, Spain

[11] CINN (CSIS-Universidad de Oviedo), El Entrego, Spain

Email: *jfullerton@anl.gov; **n.leo@lboro.ac.uk; ***amalio.fernandez-pacheco@tuwien.ac.at

**Abstract**

Magnetic skyrmions are chiral spin textures with rich physics and great potential for unconventional computing. Typically, skyrmions form in bulk crystals with reduced symmetry or ultrathin film multilayers involving heavy metals. Here, we demonstrate the formation of fractional Bloch skyrmion tubes at room temperature by 3D printing ferromagnetic double-helix nanowires with two regions of opposite chirality. Using X-ray microscopy and micromagnetic simulations, we show that the coexistence of vortex and anti-parallel spin states induces the formation of fractional skyrmion tubes at zero magnetic fields, minimising the energy cost of breaking the coupling between geometric and magnetic chirality. We also demonstrate control over zero-field states, including pure vortex, or mixed skyrmion-vortex states, highlighting the magnetic reconfigurability of these 3D nanowires. This work shows how interfacing chiral geometries at the nanoscale can enable advanced forms of topological spintronics.

**Introduction**

The control of magnetic topology and emergent behaviour is key for developing next generation spintronic devices and for unconventional computing applications [1-6]. Central to this, we find topological states such as the magnetic skyrmion, with attractive properties for computing, including topological protection, low energy motion under spin-polarised currents, and deterministic or stochastic operational regimes [7-10]. Skyrmions feature a whirling spin configuration with a well-defined chirality (or handedness), and are characterised by the topological charge, $Q$ [11]:

$$Q = \frac{1}{4\pi}\int \boldsymbol{M} \cdot \left(\frac{\partial \boldsymbol{M}}{\partial x} \times \frac{\partial \boldsymbol{M}}{\partial y}\right) dx dy$$

describing how much of the magnetisation, $M$, wraps around the unit sphere [11].

Typically, a combination of symmetry breaking and high spin-orbit coupling can result in the Dzyaloshinskii-Moriya interaction (DMI), an antisymmetric exchange energy which favours a given sense of spin rotation and allows skyrmion formation. Bulk DMI can be present in noncentrosymmetric crystals such as B20 materials, or interfacial DMI can be induced in thin film multilayers formed of ferromagnets and heavy metals [12-14]. Extrinsic parameters such as temperature, applied magnetic fields and electrical currents are often required to nucleate and stabilise individual skyrmions or skyrmion lattices [15-17].

Beyond skyrmions, the expansion of nanomagnetism into 3D brings great opportunities for creating spintronic devices utilising more complex topological magnetic textures. Whereas skyrmions are normally considered a 2D spin texture, recent works in crystals and extended multilayers focus on 3D chiral spin textures. This includes skyrmion tubes and braids, cocoons, chiral bobbers, Bloch points and hopfions [18-26]. Furthermore, 3D nano-patterning is an effective technique to control magnetisation through geometric curvature and confinement [27-32]. We have previously shown how 3D-printed helical nanowires can control the chirality of magnetic vortices and domains, and domain wall coupling in-between neighbouring helices can create topological stray fields [28-29].

In this work, we demonstrate the room temperature formation of fractional Bloch skyrmion tubes in 3D cobalt double-helices with two interfaced chiral regions. The samples are prepared via advanced 3D nano-printing and characterised by X-ray magnetic microscopy. By carefully tuning the 3D geometry, we can favour the coexistence of vortex and anti-parallel magnetic states in a helical nanowire. This coexistence, in combination with a magnetic field protocol, enables the formation of a fractional skyrmion at zero fields, where the magnetochirality no longer follows the geometric chirality. Micromagnetic simulations confirm this scenario, with an enhancement of the overall topological charge in the region where the magnetochirality opposes the geometric chirality and the skyrmion is formed. This work demonstrates chiral nano-patterning as a powerful platform to form complex topological spin states in 3D magnetic nanowires.

**Results**

Our work focuses on a 3D chiral cobalt nanowire (Fig. 1a) formed of intertwined helices strongly coupled via exchange and dipolar interactions [28]. The geometric chirality, $\chi_G$, of the helices is reversed approximately halfway through its length, resulting in two regions of opposite $\chi_G$ (right-handed, RH, at the bottom, and left handed, LH, at the top), joined by a chirality interface (marked with * in all figures). By using focused electron beam induced deposition (FEBID), we can experimentally realise these nanowires with feature sizes comparable to characteristic magnetic lengthscales [33] (Fig. 1b). Previously, we identified two possible remanent spin states for a helical chiral region [28]. The first one is the vortex state (left panel in Fig. 1c), where the magnetisation in each helical strand has the same axial polarity ($P$) and the chiral geometry defines the resulting outer circulation ($C$) [28]. The second remanent state observed is the anti-parallel (AP) state (right panel in Fig. 1c). Here, the two strands adopt an opposite magnetic polarity, resulting in helical domains separated by a chiral Bloch domain wall [28]. Importantly, for both remanent states the magnetochirality, $\chi_M$, has been found to be governed by the patterned geometrical chirality [28], a condition we denote here as $\chi_M = \chi_G$. Consequently, as our nanowire (Fig. 1b) combines two regions of opposite geometrical chirality, we promote the interfacing of spin states with opposite magnetochirality [28, 34-35].

In addition to controlling magnetochiral effects, 3D nanopatterning can tune the magnetic energy landscape via geometric control. Here, we establish how the helical pitch (describing local curvature [28-31]) and strand separation (determining the interaction between strands [28, 36]) define the magnetic ground state. Fig. 1d shows a micromagnetic phase diagram describing where the vortex (orange) and AP (blue) states are the energetically favourable state at zero field. The diagram is computed for a nanowire diameter of 86 nm and a saturation magnetisation of 900 kA/m, corresponding to our experimental system with typical values for FEBID cobalt [28, 36-37]. The ground state for a pitch below 250 nm is the vortex, which is also moderately favoured by decreasing strand separation. For small pitch and separation, the helical features of the nanowire start to fade away, with the system tending towards a thick achiral nanowire. Conversely, the AP state is energetically favourable at high pitch and separation, with a clearer distinction of the two strands, and where edge effects become more relevant. For this work, we choose a pitch of 200 nm and a separation of 66 nm (marked by a cross in Fig. 1d). As a result, the chiral regions of our nanowire favours vortex textures, while remaining close to degeneracy with the AP state. Consequently, we promote a coexistence of states, aiding the formation of hybrid metastable spin configurations under applied fields, as shown later.

To investigate the magnetic field-driven evolution of this type of chiral nanowire, we use soft X-ray ptychography combined with X-ray magnetic circular dichroism (XMCD). Ptychographic reconstructions of overlapping scattering patterns allows imaging of the magnetisation projected along the X-ray beam (Fig. 2a) [38-42]. The nanowire is mounted with a slight tilt away from the X-ray beam, allowing for sensitivity to in-plane ($x$-axis) and axial ($z$-axis) magnetic components as both are projected along the beam direction. We study the reversal process of the nanowire under uniaxial magnetic fields up to ±180 mT along the $z$-direction.

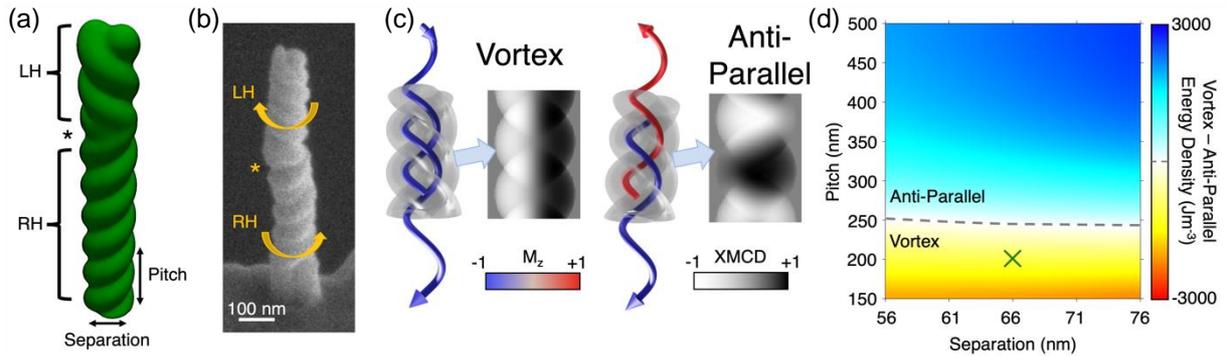

**Figure 1: A three-dimensional chiral nanowire and its characteristic remanent magnetic states. (a)** Simulated model consisting of a left-handed (LH) double helix at the top and a right-handed (RH) double helix at the bottom (RH). The two chiral regions are connected by a chirality interface (*). Helix pitch and strand separation are indicated as geometrical parameters. **(b)** SEM image of an experimentally realised 3D printed cobalt double helix (image at 45° stage tilt). **(c)** Schematic depictions of the remanent vortex (left) and anti-parallel (right) states for a LH intertwined double-helix nanowire, coloured by the $M_z$ component. The expected X-ray magnetic contrast calculated from micromagnetic simulations is shown on the right of both states. **(d)** Phase diagram at zero field indicating for which values of pitch and separation the anti-parallel (blue) or vortex (orange) state is the ground state of a chiral region. The dashed line marks where the two states are degenerate. The cross marks the combination of parameters used for the simulated structure in (a), to match the experimental structure in (b).

We first investigate the remanent spin states present after magnetic saturation. Figures 2b and 2c show XMCD images at zero field after applying axial fields of –180 mT and +180 mT, respectively. Both images show a transverse black-white/white-black contrast across the nanowire, characteristic of a vortex state [28, 43] (see simulated contrast in Fig. 1c). We observe that the XMCD signal is opposite in the LH and RH regions, a contrast that reverses for opposite magnetic field directions. Complementary micromagnetic simulations further our understanding of the underlying spin structure. Figures 2d-f show simulations of remanent spin states after applying negative (Fig. 2e) and positive (Fig. 2f) saturating fields. Here, the axial field sets the magnetic polarity, and the circulation then conforms to the chiral geometry, creating two vortex state tubes with opposite circulation and $\chi_M = \chi_G$ for both chiral regions. Simulations reveal that at the chirality interface, the magnetisation is aligned with the polarity, resulting in a Néel-like rotation of the outer spins to mediate the reversal of the magnetic circulation. The good agreement with the experimental images indicates a strong coupling between the geometric and magnetic chirality in the nanowire, with the remanent spin configuration after an applied saturating field dictated by the geometric handedness.

To further confirm that indeed, $\chi_M = \chi_G$ is obtained experimentally, we make use of the fact that the nanowire is tilted away from the X-ray beam (~15-20°), allowing some sensitivity to the magnetic polarity. We evaluate this in Fig. 2g and 2h, by taking XMCD intensity profiles across the top and bottom helices of the nanowire for the images in Fig. 2b and 2c. The resulting intensity profiles are asymmetric and show a larger magnitude of the XMCD signal (indicated by orange dots) on the left- and right-hand sides of the top and bottom chiral regions, respectively. We consider the origin of this asymmetry in Fig. 2i by evaluating the effect of structure tilt on the XMCD signal, assuming a vortex state where $\chi_M = C \times P$ [44-46]. If the structure was directly perpendicular to the X-ray beam, we would be only sensitive to the circulation, and hence see an even transverse black-white or white-black contrast. However, a tilted structure exhibits an asymmetric XMCD signal due to the projection of the magnetic polarity along the X-ray beam. This asymmetry manifests as an increase in signal intensity on one side of the nanowire, and a reduction on the other side, as the simulated XMCD signals in Fig. 2i show. For our experimental configuration, a vortex with a LH magnetochirality favours a larger XMCD magnitude on the left part of the nanowire (top row, Fig. 2i), and conversely for a RH magnetochirality (bottom, Fig. 2i). Supplementary section 1 provides further details. From these results, we experimentally confirm that at remanence after saturation, the magnetochirality is dictated by the geometric chirality across the whole nanowire.

The asymmetry of XMCD signal discussed above also allows us to extract both magnetic polarity and circulation as a function of external magnetic fields. For this, we fit the profile to a vortex tube state in both regions (see supplementary section 1 for details). Fig. 2j shows these values for a branch of a major hysteresis loop, where the field is varied from +180 to –180 mT in steps of –20 mT (see supplementary section 2 for the full image set). For all fields, the magnetic circulation has opposite signs in the LH and RH regions (green and black lines, respectively), as expected when $\chi_M = \chi_G$ for the same polarity. The circulation reaches a maximum magnitude at –20 mT and is followed by a simultaneous reversal at –40 mT in both regions. Correspondingly, both polarities switch signs at –40 mT, therefore maintaining the condition $\chi_M = \chi_G$ in both chiral regions after magnetic reversal. To match the experimental analysis, we plot the circulation and polarity vs. applied fields for a major hysteresis loop obtained from simulations (Fig. 2k). In general, there is an excellent agreement between experiments and simulations. However, the smaller field step used in the simulations (2 mT) reveal that the switching of the circulation in the RH chiral region takes place in two steps during the magnetic reversal (–26 to –30 mT, highlighted by grey).

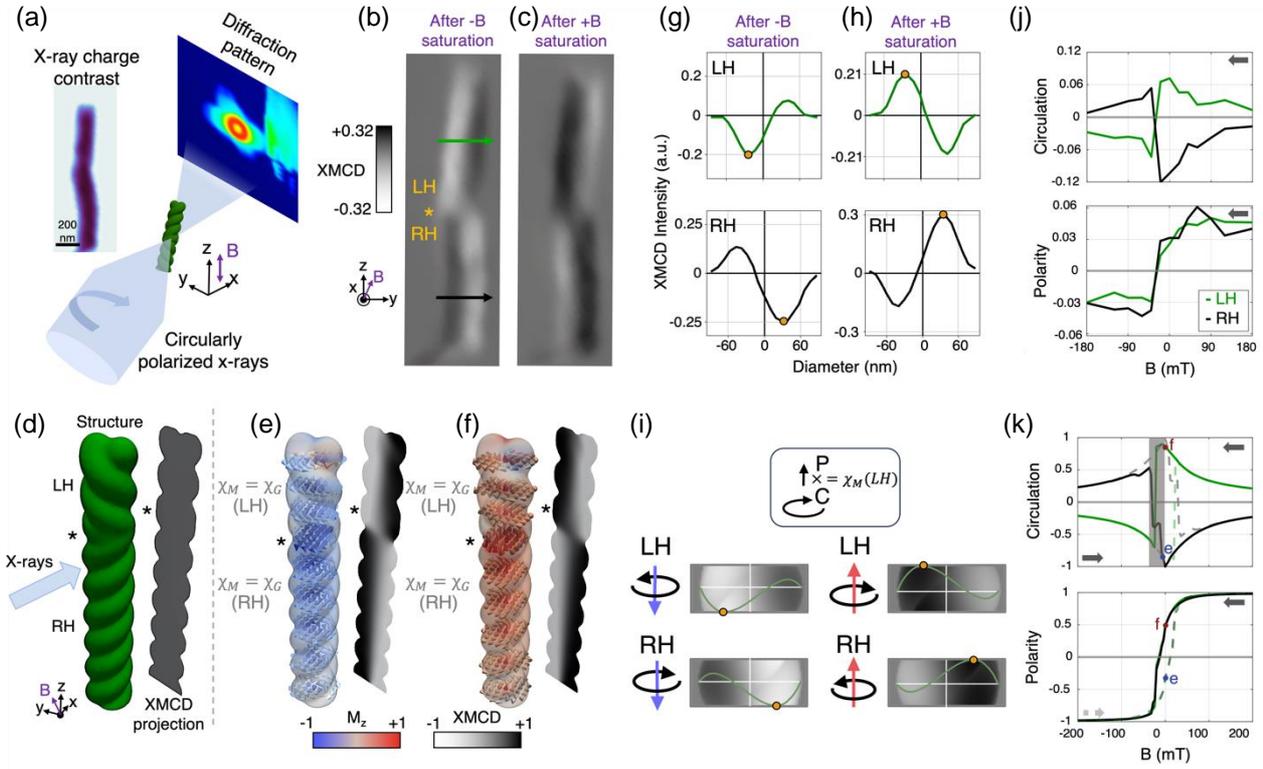

**Figure 2: Geometric-magnetic chirality coupling in an interfaced chiral nanowire.** (a) Ptychographic imaging set-up providing XMCD projections along the x-axis, while applying $B_z$ magnetic fields along the nanowire axis, with a small transverse component. The inset shows an X-ray charge contrast image, reconstructed from diffraction patterns obtained by scanning the beam over the sample. (b, c) XMCD images of the nanowire at remanence after saturation with a -180 mT (b) and a +180 mT (c) magnetic field, with vortex states of opposite magnetochirality in both chiral regions. (d) Simulation model to match the experimental setup, including the nanowire and the XMCD image obtained in transmission. (e, f) Simulated remanent states after applying a negative (e) and positive (f) saturating magnetic fields. The magnetisation is shown as xy cross-sections coloured by the $M_z$ component and with the corresponding projected XMCD contrast to the right. (g, h) Extracted experimental XMCD line profiles for the images in (b) and (c), respectively. The intensity profiles correspond to a circulating vortex, with the core leading to an asymmetry in the signal between left and right parts of the nanowire, as marked by the orange dot. (i) Schematic depiction and computed XMCD contrast on the effect of a 20° structure tilt on the XMCD signal of the states shown in (e, f). A left(/right)-handed vortex state results in an increase in intensity on the left(/right)-hand side of the nanowire. The favoured side is indicated by orange dots. The definition of the magnetochirality ($\chi_M$) for a vortex state as the product of the central polarity and outer circulation, is also included. (j) Fitted values of circulation and polarity for each chiral region (green: LH, black: RH) for a magnetic field cycle (from +180 to -180 mT in steps of -20 mT). The geometric-magnetic chirality coupling is preserved across the cycle for all field values recorded. (k) Circulation and polarity obtained from a simulated hysteresis loop between fields of ±200 mT in steps of 2 mT, showing a good agreement with experiments. The reverse branch of the loop is displayed with dashed lines. The blue and red dots and corresponding + and - symbols indicate the values associated with the states at remanence plotted in (e, f). The grey region corresponds to an intermediate state during switching, observed during the simulated loop when a small field step is employed.

We experimentally investigate the existence and nature of this intermediate state predicted by simulations by imaging the nanowire under smaller field steps. Fig. 3a (i-vi) show selected XMCD images of a major loop between fields of 0 and 180 mT, starting at remanence after negative saturation. With increasing positive fields, a distinct intermediate state forms (iii). Here, the top LH region switches to the opposite vortex state *i.e.*, reversing both polarity and circulation, therefore maintaining the $\chi_M = \chi_G$ condition. Simultaneously, an AP state forms in the bottom RH region, identified in the XMCD images by alternating vertical black and white bands [28] (see the simulated contrast on the right panel of Fig. 1c). Finally, the magnetisation fully reverses, and the opposite vortex state is obtained (v), before saturating at higher fields (vi). These results thus show that the magnetic reversal of the nanowire takes place via a hybrid AP-vortex state. We understand this finding to be a direct consequence of the geometric control discussed for Fig. 1d, where we promote co-existence of the two states by appropriate tuning of the double-helix parameters.

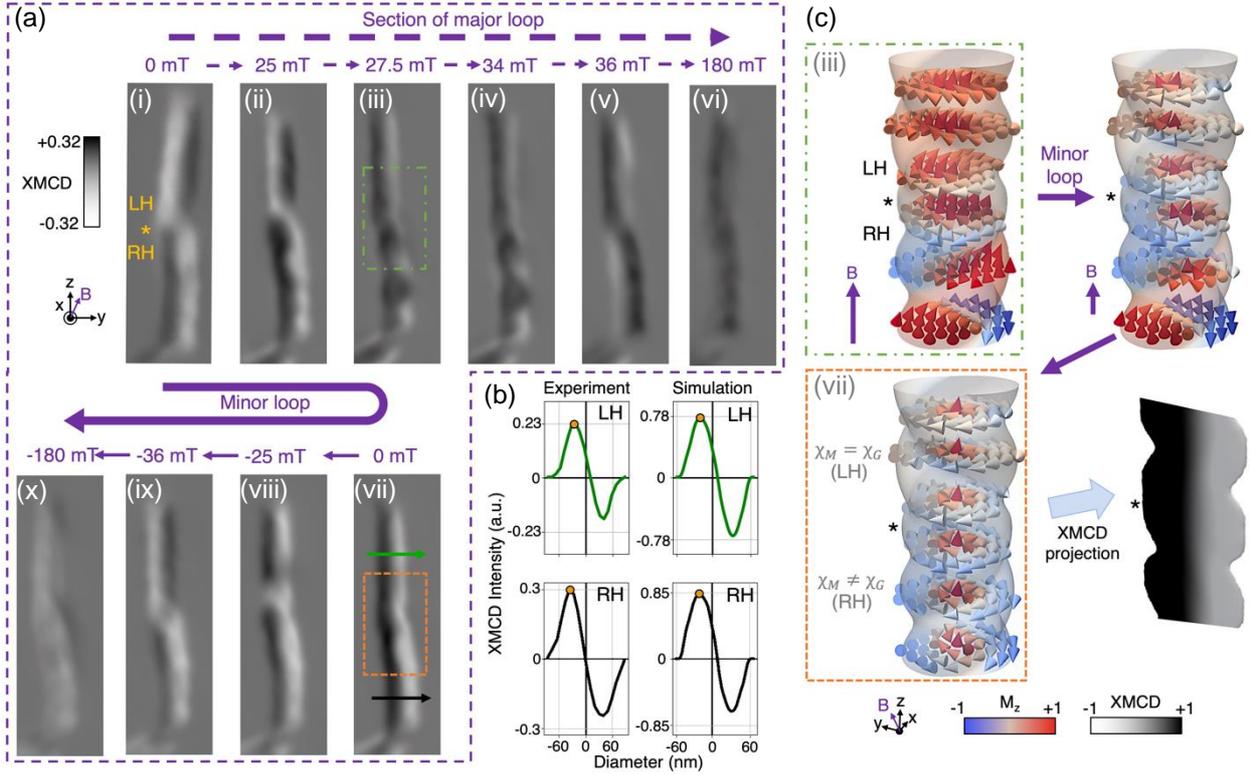

**Figure 3: Breaking of geometric-magnetic chirality coupling, leading to the formation of a skyrmion tube at remanence. (a)** Selected XMCD images of the nanowire during a major (i-vi) and a minor (iii, vii-x) hysteresis loop. A second remanent state is formed during the minor loop sequence (vii), featuring the same magnetic circulation in both chiral regions. **(b)** Comparison between experimental (left) and simulated (right) XMCD line profiles. The experimental line profile is extracted from the second remanent state (a, vii), see green and black arrows, whereas the simulated line profile is extracted from the simulations shown in (c). In both cases, the intensity profile corresponds to a vortex-like state in both chiral regions with the same circulation and polarity, as indicated by the higher intensity on the left side, marked by orange dots. The same magnetochirality is thus present in both, despite an opposite geometric chirality. **(c)** Snapshots from micromagnetic simulations reproducing the minor loop field sequence as in experiments. The region shown corresponds to the area around the chirality interface, as the one indicated by the green and orange boxes in the images iii and vii, respectively, of part (a). The magnetisation is represented by multiple $xy$ cross-sections coloured by the $M_z$ component, with the first and last images corresponding to the (iii) and (vii) states in (a). Image (iii) shows a hybrid vortex/AP state, and (vi) reveals the spin configuration of the second remanent state, with a uniform core polarity and circulation throughout the full nanowire due to the presence of a skyrmion stube formed at the bottom. The XMCD signal of this state is shown on the right, in good agreement with experiments.

To investigate the implications of forming this hybrid AP-vortex state, we perform a minor loop initiated from this intermediate state (Fig. 3a, iii and vii-x). When the magnetic field is removed, a second distinct remanent state forms (vi). Here, the AP state in the RH region relaxes into a spin configuration with a very similar XMCD contrast as the LH region, thereby creating a transverse black-white XMCD signal spanning the entire nanowire. To assess the magnetic polarity of state vii, we do as before, *i.e.*, take line profiles of the XMCD signal across both chiral regions (Fig. 3b, left). Both line profiles show a larger intensity on the left of the nanowire in both chiral regions, indicating a uniform left-handed magnetochirality throughout the whole nanowire. Consequently, these results imply that $\chi_M \neq \chi_G$ in the bottom chiral region. Subsequent application of negative fields during this minor loop creates a region of local AP alignment around the chirality interface (viii), before the original two-vortex state is recovered at −36 mT (ix), where $\chi_M = \chi_G$ for both regions again. Hence, these experiments indicate the ability to locally break and recover the geometric-magnetic chiral coupling in interfaced 3D chiral nanowires under minor loop sequences.

Fig. 3c shows micromagnetic simulations to describe the formation mechanism and magnetic configuration of this second remanent state. We start from a hybrid AP-vortex state with a positive applied field, matching the intermediate state in experiments (green box in iii). Overall, there is a large $+M_z$ component throughout the nanowire, due to the applied field, with a small $-M_z$ component in the bottom AP section. As the applied field is reduced (follow the purple arrows, Fig. 3c), the magnetic coupling between the bottom and top regions, mediated by the chirality interface, promotes a vortex-like state also in the bottom region. In this region, the previously $+M_z$ of the AP domain forms its central polarity, and the $-M_z$ domain forms its circulating shell. Importantly, the bottom region evolves from an AP state to a state with a significantly larger outer axial magnetisation than a typical vortex. Here, the reversed $+M_z$ core (red) is surrounded by a shell with a significant opposite $-M_z$ (blue), thereby resembling a skyrmion tube [47-49].

We investigate further the spin configuration of this skyrmion state, comparing it with the vortex tube normally observed (favoured by geometry). Either a favourable or opposing geometric-magnetic chirality in both cases significantly impacts the magnetic topology, as well as the behaviour of both spin states under applied fields. Figures 4a and 4b show a visual, micromagnetic comparison between the vortex and skyrmion tubes obtained from simulations, at their minimum and maximum topological charge ($Q$), respectively, as described below. We first evaluate the difference in topology for the two distinct remanent states observed: the two-vortex state (Fig. 4c), where $\chi_M = \chi_G$ for the whole nanowire; and the hybrid vortex/skyrmion state (Fig. 4e), where $\chi_M = \chi_G$ for the top chiral region and $\chi_M \neq \chi_G$ for the bottom one. Figures 4d and 4f show the field dependence of $Q$ evaluated for each $xy$ plane along the length of the nanowire, obtained from micromagnetic simulations. The magnetic field is applied against the axial core direction, and the line colours (red-green-purple) correspond to magnetic field values increasing from zero to just before the reversal of the magnetisation takes place (further simulation snapshots and discussion are contained in supplementary section 5). As a reference, for each 2D plane, $Q = 0$ would correspond to a uniformly magnetised state, $Q = 0.5$ to a planar vortex, and $Q = 1$ to a skyrmion. For skyrmion-based textures, $Q$ is often assumed to take integer values.

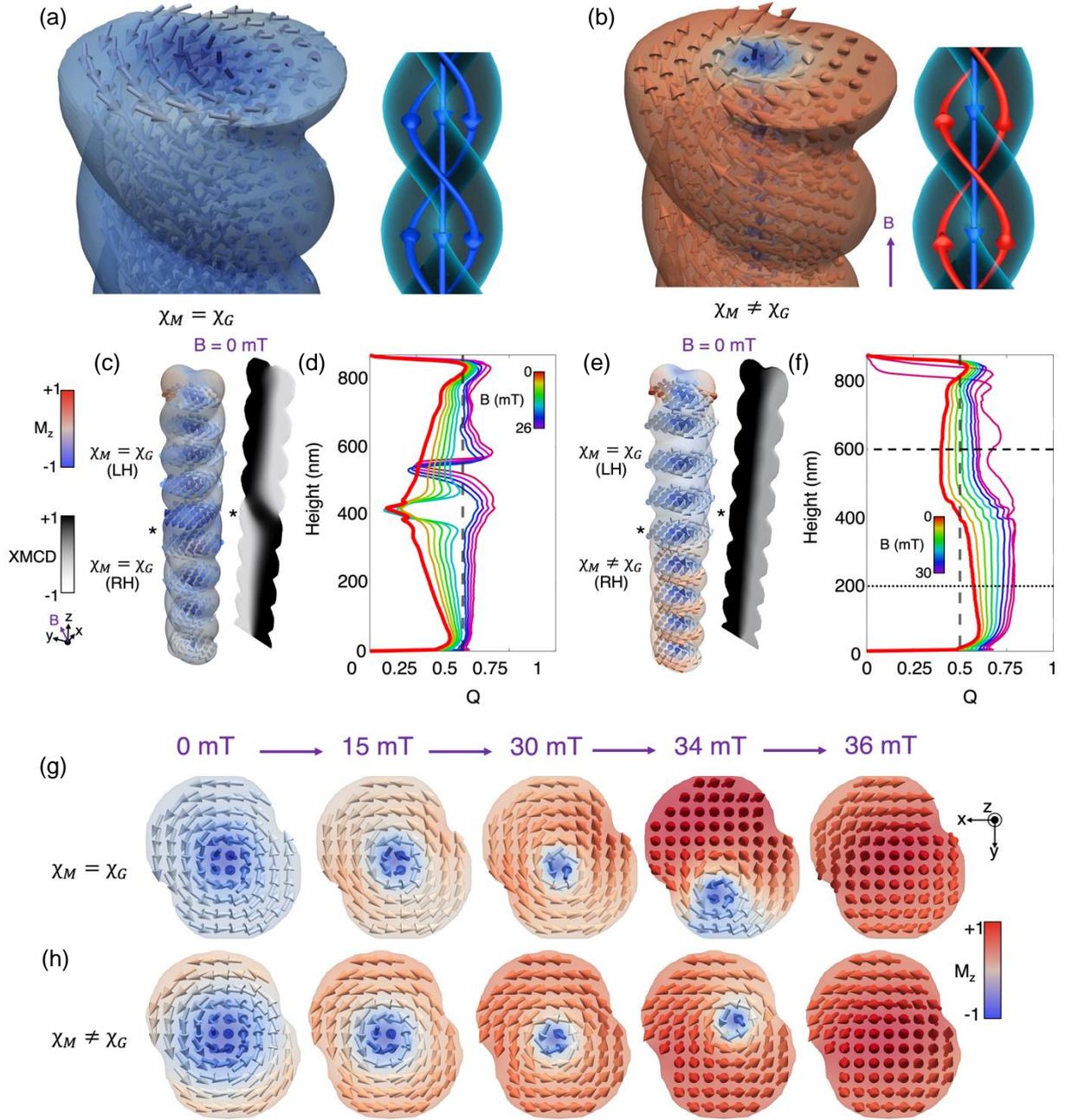

**Figure 4: A topological comparison between vortex and skyrmion states, at remanence and under applied fields.** (**a, b**) 3D micromagnetic simulation and schematic of a vortex and a skyrmion tube, respectively, obtained in chiral nanowires due to 3D geometric effects. (**c**) Micromagnetic simulations coloured by the $M_z$ component and expected XMCD contrast of the two-vortex state. (**d**) Calculated topological charge ($Q$) in multiple xy planes vs. height (z coordinate) with increasing applied field (red to green to blue) for the two-vortex state. (**e**) Micromagnetic simulations coloured by the $M_z$ component and expected XMCD contrast of the vortex/skyrmion state. The transition from one vortex to the other takes place by a dip in topological charge, which is localised at the chirality interface for lower field values. (**f**) Calculated topological charge ($Q$) in multiple xy planes vs. height (z coordinate) with increasing applied field (red to green to blue) for the vortex/skyrmion hybrid state. The dashed and dotted lines correspond to the cross-sections in (g) and (h). The breaking of geometric-magnetic chirality coupling in the bottom chiral region results in a more continuous transition of $Q$ from one chiral region to the other, with larger values at the bottom corresponding to a fractional skyrmion. (**g, h**) Micromagnetic evolution with applied field of spin textures in an xy plane of the vortex/skyrmion state. At the top (vortex), $\chi_M = \chi_G$; at the bottom, (fractional skyrmion of Bloch type) $\chi_M \neq \chi_G$.

boundary conditions, confined geometries may result in skyrmionic spin textures with non-integer values of $Q$, including fractional/incomplete skyrmions [50-54].

We consider first the two-vortex state (Fig. 4d). At zero field (red line) a maximum value of $Q \approx 0.4$ is found at the nanowire ends. This is smaller than for a planar vortex, as the shape anisotropy created by a helical geometry favours a more axial configuration. Over the length of the nanowire, $Q$ progressively decreases to $Q \approx 0.1$ at the chirality interface, where the reversal of magnetochirality has most spins aligned axially. When a field is applied, $Q$ increases throughout the whole structure. Local areas with $Q \approx 0.6$ become present close to the switching field as spins align along the field direction. However, a local minimum in $Q$ remains throughout the field protocol. While the origin of this minimum at lower fields is due to the axial alignment of spins at the chirality interface, its presence at higher fields is a result of the nucleation and subsequent displacement of an AP state from the chirality interface.

Conversely, in the vortex/skyrmion state (Fig. 4f), $Q$ is generally larger and more uniform than for the previous case. Specifically, as there is no reversal in magnetochirality between the $\chi_M = \chi_G$ (top, LH) and $\chi_M \neq \chi_G$ (bottom, RH) regions, we observe $Q$ values close to 0.4 and 0.6 at remanence, respectively. A larger value of $Q$ in the bottom region is directly linked to the breaking of chirality coupling via the formation of a Bloch skyrmion, as $xy$ cross-sections of the magnetisation as a function of field confirm (see Figs. 4g and h). Here, the spin circulation in the outer shell is opposite to the geometrical slope of the helical strands, resulting in an increase of $+M_z$ component to minimise surface charges, hence covering a larger solid angle. The skyrmion state observed is a fractional skyrmion, with $Q$ values always smaller than 1, and which are further enhanced with increasing field, reaching a maximum value of 0.8 at fields close to switching. This second remanent state shows how by reconfiguring the magnetisation to oppose the geometrically imprinted chirality, we can access a larger range of fractional topological charges. Specifically, we observe a continuous range of $0.4 < Q < 0.8$ over the nanowire and field range, with the maximum angle of the outer spins limited by the slope of the helical strands.

**Discussion**

In this work, we demonstrate a mechanism to form fractional skyrmions by means of 3D geometrical effects, based on interfacing helical nanowires of opposite chirality. While the creation of this skyrmion state with $\chi_M \neq \chi_G$ is costly in both exchange and dipolar energy, its formation is facilitated by several factors. Firstly, by designing a nanowire where the vortex and AP states are close to degeneracy (Fig. 1d), we promote the formation of hybrid AP-vortex states during magnetic reversal (Fig. 3c, iii). When the field is removed, the presence of opposite axial AP domains help initialise the skyrmion, with one domain forming the core, and the other forming the outer shell. Additional experiments in supplementary section 3 show how non-optimised material/geometric parameters prevent the formation of AP states, and therefore the emergence of skyrmions via a minor loop. Secondly, interfacing opposite geometric chiral regions produces a strong magnetic coupling between them. A vortex in one region, interfaced with a metastable AP in the other, will in principle favour the formation of a uniform vortex tube with same polarity and circulation throughout the whole nanowire, as the field is ramped down to zero. However, a vortex with opposite magnetochirality to that favoured by the chiral geometry is energetically too costly. Hence, a fractional skyrmion with a larger topological charge (*i.e.*, an axial outer spin configuration) is formed, minimising the creation of magnetic surface charges. This skyrmion tube still opposes the geometric chirality, but the energy cost is lower than for a vortex tube. In supplementary information section 4, we discuss energetic comparisons between potential uniform circulation magnetic states, finding that under these conditions a fractional skyrmion is preferred. The formation of different skyrmion phases in interfaced ferro/ferri magnetic thin films has been recently demonstrated [19]. Here, we realise instead a purely 3D geometrical mechanism based on interfacing helical nanowires of opposite geometric chirality. Interestingly, the access to a regime where geometric and magnetochirality oppose to each other, directly results in the formation of spin states with higher topological numbers.

**Conclusions**

Magnetic nanowire devices are currently being intensively studied for their potential in non-conventional computing applications, such as the racetrack memory, neuromorphic computing and magnonics[55-56]. The controlled manipulation of topological spin states, such as skyrmions within these nanowires, is anticipated to pave the way for innovative, energy-efficient computing technologies. While most studies have focused on planar nanowires due to lithographic limitations, expanding these structures into 3D opens new possibilities in spintronics, including enhanced storage capacity and device interconnectivity, as well as enabling novel geometrically-driven magnetic effects.

In this work, we demonstrate that interfacing nanowires with opposite geometric chirality results in reconfigurable systems, where vortex and fractional skyrmion tubes may coexist at zero magnetic field and room temperature. The emergence and coexistence of high-order topological spin states is a direct consequence of chiral patterning, offering new possibilities for reservoir computing [57-58] and the formation of complex 3D spin textures like hopfions [23-24], in nanowire-based architectures.

**Methods**
*Nanofabrication*

Magnetic Cobalt double helices were grown on Cu TEM grids (pre-cut with a focused ion beam) using focused-electron-beam-induced deposition (FEBID) from a $Co_2(CO)_8$ precursor with a Thermo Fisher HELIOS-600 Dual beam SEM + FIB. Initial 3D computer models of the nanowires were transformed to stream files using our open-access code f3ast [34]. Accelerating voltages 5 kV and beam current 43 pA at a gas pressure of $3 \times 10^{-6}$ mbar was used with an initial growth rate (of $GR_0 = $ 40-60 nms$^{-1}$), and a variable dwell time to account for thermal effects during the nanowire growth [33].

*X-ray magnetic ptychography*

X-ray microscopy experiments were performed at the HERMES beamline at SOLEIL synchrotron, France. To obtain magnetic contrast, an energy tuned to the Co L3 edge at 782 eV was used. The sample was mounted in a magnetic field holder, allowing to apply fields up to $\pm 180$ mT both along the direction of the beam (*i.e.*, x-axis in Fig. 2a) and approximately along the long axis of the structure (z-axis in Fig. 2a). To obtain the ptychographic images, the amplitude and phase of both the object (*i.e.*, sample) and source (*i.e.*, X-ray beam) were reconstructed from the set of diffraction patterns using a GPU-accelerated reconstruction algorithm provided in the PyNX package. The magnetic spatial resolution of this technique is the result of the diffraction of the beam with the object, in this case 8 nm/pixel [38-42]. XMCD contrast images were obtained from the absolute value of the complex-valued reconstructed object image, after image registration (with canny edge filter) and background normalisation. Extraction of line profiles and subsequent fitting routine to extract the amplitudes for circulation and polarity components are described in Supplemental Material Section 1.

*Micromagnetic simulations*

Micromagnetic simulations were performed using MuMax3 [59]. A saturation magnetisation $M_S = 900$ kAm$^{-1}$, exchange stiffness $A = 10^{-11}$ Jm$^{-1}$, and zero magneto-crystalline anisotropy for a cell size of 2 nm was used to represent the experimental FEBID structures (see Supplementary for other parameters for comparison).

To map the double helix's energy landscape as a function of geometry (Fig. 1d), the pitch and separation of the intertwined helices were varied between 150-500 nm and 56-76 nm, respectively, for wires with an 86 nm diameter. In each case, the magnetisation was initialised as either a uniform vortex tube (with magnetochirality matching the geometry), to form the vortex state, or with the two helical strands with opposing $M_z$ components, to form the anti-parallel state. These initial configurations were then fully relaxed to compare their relative energy density. A wire separation of 66 nm and pitch of 200 nm was chosen for the subsequent simulations to support the experimental work. The simulated XMCD projections were calculated as the projection of the magnetisation vectors with the defined direction of the X-ray beam.

To compare the topology of the two-vortex and hybrid vortex/fractional skyrmion states, each state was initialised respectively as either a uniform axial or vortex configuration across the whole structure (*i.e.*, both chiral regions) before relaxing into their respective remanent states. The topological charge $Q$ was then calculated for each xy slice of magnetisation along the z-axis using the equation:

$$Q = \frac{1}{4\pi}\int \boldsymbol{M} \cdot \left(\frac{\partial \boldsymbol{M}}{\partial x} \times \frac{\partial \boldsymbol{M}}{\partial y}\right) dx\, dy$$

where $\boldsymbol{M}$, the magnetisation, is integrated over a 2D slice.

**Acknowledgements**
This work was supported by the EPSRC and the Centre for Doctoral Training (CDT) in Photonic Integration and Advanced Data Storage (PIADS), RCUK Grant No. EP/L015323/1, the European Community under the Horizon 2020 Program, Contract No. 101001290 (3DNANOMAG). JF would also like to acknowledge the U.S. Department of Energy, Office of Science, Office of Basic Energy Sciences, Materials Sciences and Engineering Division. The work of NL was also supported by a UKRI Future Leader Fellowship [grant number MR/X033910/1 (LIONESS)]. LS acknowledges support from the EPSRC Cambridge NanoDTC EP/L015978/1. A.H.-R. acknowledges the support from Spanish MCIN/AEI/10.13039/501100011033/FEDER, UE under grant PID2022-136784NB. We acknowledge Synchrotron SOLEIL for providing the synchrotron radiation facilities (Proposal No. 20191674). We would also like to thank Sam McFadzean and William Smith at the Kelvin Nanofabrication centre at the University of Glasgow for technical support.